\documentclass[aps,prc,twocolumn,showpacs,showkeys,superscriptaddress,amsmath,amssymb,nofootinbib]{revtex4}

\usepackage{graphicx}
\usepackage{dcolumn}
\usepackage{bm}
\begin{document}


\title{Gamow-Teller Unit Cross Sections for ($t$,$^{3}$He) and ($^{3}$He,$t$) Reactions}

\author{G. Perdikakis}
\email{perdikak@nscl.msu.edu}\affiliation{National Superconducting Cyclotron Laboratory, Michigan State University, East Lansing, MI 48824-1321, USA} \affiliation{Joint Institute for Nuclear Astrophysics, Michigan State University, East Lansing, MI 48824, USA}
\author{R.G.T. Zegers}
\email{zegers@nscl.msu.edu}
\affiliation{National Superconducting Cyclotron Laboratory, Michigan
State University, East Lansing, MI 48824-1321, USA}
\affiliation{Joint Institute for
Nuclear Astrophysics, Michigan State University, East Lansing, MI 48824, USA}
\affiliation{Department of Physics and
Astronomy, Michigan State University, East Lansing, MI 48824, USA}
\author{Sam M. Austin}
\affiliation{National Superconducting Cyclotron Laboratory, Michigan State University, East Lansing, MI 48824-1321, USA}
\affiliation{Joint Institute for Nuclear Astrophysics, Michigan State University, East Lansing, MI 48824, USA}
\author{D. Bazin}
\affiliation{National Superconducting Cyclotron Laboratory, Michigan State University, East
Lansing, MI 48824-1321, USA}
\author{C. Caesar}
\affiliation{Joint Institute for
Nuclear Astrophysics, Michigan State University, East Lansing, MI 48824, USA}
\affiliation{Johannes Gutenberg Universit\"{a}t, Mainz, Germany}
\author{J.M. Deaven}
\affiliation{National Superconducting Cyclotron Laboratory, Michigan State University, East
Lansing, MI 48824-1321, USA} \affiliation{Joint Institute for Nuclear Astrophysics,
Michigan State University, East Lansing, MI 48824, USA} \affiliation{Department of Physics and Astronomy, Michigan State
University, East Lansing, MI 48824, USA}
\author{A. Gade}
\affiliation{National Superconducting Cyclotron Laboratory, Michigan
State University, East Lansing, MI 48824-1321, USA}
\affiliation{Department of Physics and
Astronomy, Michigan State University, East Lansing, MI 48824, USA}
\author{D. Galaviz}
\altaffiliation[present address: ]{Centro de Fisica Nuclear da Universidade de Lisboa, 1649-003, Lisbon, Portugal}
\affiliation{National Superconducting Cyclotron Laboratory, Michigan
State University, East Lansing, MI 48824-1321, USA}
\affiliation{Joint Institute for
Nuclear Astrophysics, Michigan State University, East Lansing, MI 48824, USA}
\author{G. Grinyer}
\altaffiliation[present address: ]{GANIL, BP 55027, F-14076 Caen Cedex 5, France}
\affiliation{National Superconducting Cyclotron Laboratory, Michigan
State University, East Lansing, MI 48824-1321, USA}
\author{C. J. Guess}
\affiliation{National Superconducting Cyclotron Laboratory, Michigan State University, East
Lansing, MI 48824-1321, USA} \affiliation{Joint Institute for Nuclear Astrophysics,
Michigan State University, East Lansing, MI 48824, USA} \affiliation{Department of Physics and Astronomy, Michigan State
University, East Lansing, MI 48824, USA}
\author{C. Herlitzius}
\affiliation{Joint Institute for
Nuclear Astrophysics, Michigan State University, East Lansing, MI 48824, USA}
\affiliation{Johannes Gutenberg Universit\"{a}t, Mainz, Germany}
\author{G.W. Hitt}
\affiliation{National Superconducting Cyclotron Laboratory, Michigan State University, East
Lansing, MI 48824-1321, USA}
\affiliation{Joint Institute for Nuclear Astrophysics,
Michigan State University, East Lansing, MI 48824, USA}
\affiliation{Department of Physics and Astronomy, Michigan State
University, East Lansing, MI 48824, USA}
\author{M.E. Howard}
\affiliation{Joint Institute for Nuclear Astrophysics,
Michigan State University, East Lansing, MI 48824, USA}
\affiliation{Department of Physics, The Ohio State University, Ohio 43210, USA}
\author{R. Meharchand}
\affiliation{National Superconducting Cyclotron Laboratory, Michigan State University, East
Lansing, MI 48824-1321, USA} \affiliation{Joint Institute for Nuclear Astrophysics,
Michigan State University, East Lansing, MI 48824, USA} \affiliation{Department of Physics and Astronomy, Michigan State
University, East Lansing, MI 48824, USA}
\author{S. Noji}
\affiliation{Department of Physics, The University of Tokyo, Bunkyo, Tokyo 113-0033, Japan}
\author{H. Sakai}
\affiliation{Department of Physics, The University of Tokyo, Bunkyo, Tokyo 113-0033, Japan}
\author{Y. Shimbara}
\affiliation{Graduate School of Science and Technology, Niigata University, Niigata 950-2181, Japan}
\author{E.E. Smith}
\affiliation{Joint Institute for Nuclear Astrophysics,
Michigan State University, East Lansing, MI 48824, USA}
\affiliation{Department of Physics, The Ohio State University, Ohio 43210, USA}
\author{C. Tur}
\affiliation{National Superconducting Cyclotron Laboratory, Michigan
State University, East Lansing, MI 48824-1321, USA}
\affiliation{Joint Institute for
Nuclear Astrophysics, Michigan State University, East Lansing, MI 48824, USA}
\date{\today}%

\begin{abstract}
The proportionality between differential cross sections at vanishing linear momentum transfer and Gamow-Teller transition strength, expressed in terms of the \textit{unit cross section} ($\hat{\sigma}_{GT}$) was studied as a function of target mass number for ($t$,$^{3}$He) and ($^{3}$He,$t$) reactions at 115 $A$MeV and 140 $A$MeV, respectively. Existing ($^{3}$He,$t$) and ($t$,$^{3}$He) data on targets with mass number $12\leq A\leq 120$ were complemented with new and reevaluated ($t$,$^{3}$He) data on proton, deuteron, $^{6}$Li and $^{12}$C targets. It was found that in spite of the small difference in beam energies between the two probes, the unit cross sections have a nearly identical and simple dependence on target mass number $A$, for $A\geq 12$: $\hat{\sigma}_{GT}=109/A^{0.65}$. The factorization of the unit cross sections in terms of a kinematical factor, a distortion factor and the strength of the effective spin-isospin transfer nucleus-nucleus interaction was investigated. Simple phenomenological functions depending on mass number $A$ were extracted  for the latter two. By comparison with plane and distorted-wave Born approximation calculations, it was found that the use of a short-range approximation for knock-on exchange contributions to the transition amplitude results in overestimated cross sections for reactions involving the composite ($^{3}$He,$t$) and ($t$,$^{3}$He) probes.
\end{abstract}



\maketitle
\section{Introduction}
Charge-exchange (CE) reactions at intermediate energies have been used to study spin-isospin excitations in nuclei for more than three decades \cite{OST92,HAR01}. Many of those studies have been aimed at the extraction of Gamow-Teller (GT) transition strengths. Unlike $\beta$-decay experiments, there are only weak restrictions due to the reaction $Q$-value and GT transition strengths can be extracted up to high excitation energies.

The GT transition strengths deduced from CE experiments provide stringent tests for nuclear structure calculations and serve as input for a variety of applications in which weak transition strengths play a role. Such applications include the role of electron capture and $\beta$-decay in stellar evolution (see e.g. Refs. \cite{LAN03,ADA06,ALF93,BAU05,HIT09}), neutrino nucleosynthesis (see e.g. Ref. \cite{BYE07}), constraining calculations of matrix elements for (neutrinoless) double-$\beta$ decay (see e.g. Refs. \cite{EJI05,DOH08,GRE08,YAK09}) and the response of neutrino detectors (see e.g. Refs. \cite{FUJ00,EJI98}).

A variety of charge-exchange reactions, both in the $ \Delta T_{z} =-1 $ ($\beta^{-}$) and $ \Delta T_{z} =+1 $ ($\beta^{+}$) direction have been used. Irrespective of the probe, the extraction of GT strengths from charge-exchange data is based on the proportionality between differential cross sections at vanishing linear momentum transfer ($q\approx 0$) and the strength of the corresponding GT transitions. This proportionality -- represented by the so-called `unit cross section' ($\hat{\sigma}$) -- was first studied extensively for the ($p$,$n$) reaction \cite{TAD87} and subsequently investigated for other probes and/or different beam energies  (see e.g. Refs. \cite{SAS09a,GRE04,ANN99,NAK99,FUJ96,ADA07,ZEG06}). Key for such studies is that the unit cross sections are conveniently calibrated using transitions for which the GT transition strengths ($B(GT)$) are known from $\beta$-decay $ft$ values:
\begin{equation}\label{eq:logft}
B(F)+(\frac{g_{A}}{g_{V}})^{2}B(GT)=\frac{K/g_{V}^2}{ft},
\end{equation}
where $\frac{g_{A}}{g_{V}}=-1.2694\pm 0.0028$ \cite{PDG10} and $K/g_{V}^2=6143\pm2$ s \cite{HAR09}. Here, $B(GT)$ is defined such that it equals 3 for the decay of the free neutron.
The Fermi transition strength is confined to the excitation of the Isobaric Analog State (IAS) ($J_{i}=J_{f}$ and $T_{i}=T_{f}$) in the $\Delta T_{z}=-1$ direction.

Recently, the GT (and Fermi) unit cross sections for the $ \Delta T_{z} =-1 $ ($ ^{3} $He,$t$) reaction at 140 $A$MeV have been studied for several nuclei with mass numbers ranging from 12-120 \cite{ZEG07}. A simple phenomenological relationship between the GT unit cross section and target mass number was established. It allows for the extraction of GT strengths via the ($^{3}$He,$t$) reaction for nuclei for which $\beta$-decay data are lacking for the purpose of calibrating the unit cross section. In addition, the study of Ref. \cite{ZEG07} is also important for the CE experiments that use the $\Delta T_{z}=+1$ ($t$,$^{3}$He) reaction. After several experiments that utilized a secondary triton beam produced from a primary $^{4}$He beam \cite{DAI97,DAI98,NAK00,ZEG06,COL06}, experiments are now routinely performed using tritons at 115 $A$MeV created from fast-fragmentation of $^{16}$O nuclei \cite{HIT06,HOW08,HIT09,GUE09}. In spite of the slight difference between beam energies commonly used in ($ ^{3} $He,$t$) and ($t$,$ ^{3} $He) experiments, it is expected that the reaction mechanisms of these analog probes are very similar and this was shown explicitly for the $^{26}$Mg($^{3}$He,$t$) and  $^{26}$Mg($t$,$ ^{3} $He) reactions \cite{ZEG06} and $^{13}$C($^{3}$He,$t$) and  $^{13}$C($t$,$ ^{3} $He) reactions \cite{GUE09}.

In this work, we extend the analysis of Ref. \cite{ZEG07} in two ways:
 \begin{itemize}
   \item We combine the analysis of the GT unit cross section for the ($^{3}$He,$t$) reaction at 140 $A$MeV with data obtained via the ($t$,$ ^{3} $He) reaction at 115 $A$MeV. Existing ($t$,$^{3}$He) data on $^{6}$Li \cite{NAK00}, $^{13}$C \cite{GUE09} and $^{26}$Mg \cite{ZEG06} are complemented with new results for reactions on $^{1}$H, $^{2}$H and $^{12}$C and extracted GT unit cross sections combined with those presented in Ref. \cite{ZEG07} for the ($^{3}$He,$t$) reaction.
   \item We provide a more in-depth analysis of the target mass dependence of the extracted GT unit cross sections for the ($t$,$^{3}$He) and ($^{3}$He,$t$) reactions in terms of a factorization in a kinematical, distortion, and interaction component, based on the eikonal approximation discussed in Ref. \cite{TAD87}. The evaluation is supported by calculations in the plane and distorted-wave Born approximation.
 \end{itemize}


\section{Proportionality and Extraction of GT strength}\label{sec:Theory}
In the limit of vanishing linear momentum transfer $q\approx 0$ and applying the Eikonal approximation for the effects of distortions, the differential cross section for transitions with $\Delta L=0$ excited in charge-exchange reactions at intermediate energies ($E\gtrsim 100$ $A$MeV)  can be factorized as shown in Ref. \cite{TAD87}.
For GT transitions, one finds:
\begin{equation} \label{eq:dsigma}
\left[\frac{d \sigma}{d \Omega}(q=0)\right]_{GT} =K N^{D} |J_{\sigma \tau}|^{2} B(GT).
\end{equation}
The kinematic factor $K$ is defined as
\begin{equation}\label{eq:Kfactor}
K=\frac{E_{i}/E_{f}}{(\pi\hbar^{2}c^{2})^{2}}\frac{k_{f}}{k_{i}},
\end{equation}
where $E_{i}$($E_{f}$) is the reduced energy for the incoming (outgoing) channel, and  $k_{i}$($k_{f}$) is the incoming (outgoing) linear momentum of the projectile (ejectile).
$N^{D}$ is a distortion factor and represents the influence of the mean field of the target nucleus on the incoming and outgoing scattering waves. In the limit of $q=0$:
\begin{equation}\label{eq:dist}
N^{D}=\frac{[\frac{d\sigma}{d\Omega}(q=0)]_{DWBA}}{[\frac{d\sigma}{d\Omega}(q=0)]_{PWBA}},
\end{equation}
where subscripts DWBA and PWBA refer to calculations in the distorted and plane-wave Born approximation, respectively.
$| J_{\sigma \tau} |$ is the volume integral of the central $\sigma\tau$ component of the effective interaction between nucleons in the target and projectile nuclei.

From Eq. (\ref{eq:dsigma}) the proportionality between the differential cross section at $q=0$ and $B(GT)$ is evident and the unit cross section is defined as:
\begin{equation}\label{eq:sigma_hat}
\hat{\sigma}=KN^{D}|J_{\sigma \tau}|^{2}.
\end{equation}
The boundary condition of $q=0$ for the use of Eq. (\ref{eq:dsigma}) can only be approximately satisfied in experiments. Although the differential cross section at a scattering angle of $0^{\circ}$ can be obtained from fitting the measured angular distribution at forward scattering angles to the calculated distribution in DWBA, the differential cross section is almost always associated with a reaction $Q$-value $Q=Q_{g.s.}-E_{x}\neq0$, where $Q_{g.s.}$ is the reaction $Q$ value for the transition to the ground state and $E_{x}$ the excitation energy of the residual nucleus. Therefore, an extrapolation of the differential cross section measured at finite $Q$-value to $Q=0$ is required. This is usually done by applying the following relationship:
\begin{equation}\label{eq:extrapolation}
\left[\frac{d\sigma}{d\Omega}(0,0^{\circ})\right]=\left[ \frac{\frac{d\sigma}{d\Omega}(0,0^{\circ})}{\frac{d\sigma}{d\Omega}(Q,0^{\circ})}\right]_{T} \left[\frac{d \sigma}{d\Omega} (Q, 0^{\circ})\right]_{E},
\end{equation}
where the subscript `T' refers to calculated cross sections in DWBA and the subscript `E' refers to the experimental cross section. Eq. (\ref{eq:extrapolation}) does not take into account exactly the effects of the Coulomb potential in the scattering process. The Coulomb potential causes a deceleration of the projectile and acceleration of the ejectile in the field of the target nucleus, if either is, or both are charged. Since by definition the charges of projectiles and ejectiles are different in charge-exchange reactions, the linear momentum transfer at the interaction point is slightly different from the value calculated using the initial and final momenta of the projectile and ejectile, respectively. If $Q_{g.s.}=0$, it causes the differential cross section to peak at finite negative $Q$ (the $Q$-value for which $q=0$ at the interaction point) for ($p$,$n$)-type CE reactions, and at finite positive $Q$ for ($n$,$p$)-type CE reactions. This effect is usually ignored in analyses of charge-exchange data and the momenta of the projectile and the ejectile at large distances from the interaction point are used. For the sake of consistency, we will do the same in this work, both in the extraction of cross sections at $q=0$ from the data and in the theoretical calculations.

In CE reactions, the  $\Delta J=\Delta L + \Delta S=0+1=1$ GT transition is accompanied by transitions with $\Delta J=\Delta L + \Delta S=2+1=1$. Incoherent $\Delta L=2$ contributions to the cross section can be removed prior to applying Eq. (\ref{eq:dsigma}), based on the analysis of the experimental angular distribution. Since the angular distributions associated with the $\Delta L=0$ and incoherent $\Delta L=2$ contributions are dissimilar, this is usually and reliably accomplished by performing a multipole decomposition of the measured angular distribution based on theoretically calculated angular distributions for each of the multipole components (see also Section \ref{sec:newdata}). The coherent contribution, which is largely due to the effects of the non-central tensor interaction, is a bigger source of uncertainty \cite{COL06,ZEG06,FUJ07,ZEG08}. It cannot easily be accounted for based on the experimental data as it has little effect on the angular distributions at forward scattering angles. The interference effects of the tensor force are relatively stronger for weaker GT transitions and can only be estimated by comparing theoretical reaction calculations with and without the tensor interaction included (see e.g. Refs. \cite{ZEG06, HIT09}). Of the charge-exchange reactions discussed in this paper, the extraction of the unit cross section for the $^{58}$Ni($^{3}$He,$t$) reaction was shown to be significantly affected (20\%) by interference from the tensor interaction, as discussed in detail in Ref. \cite{COL06}.

Cross section calculations in the present work are performed in the Distorted-Wave Born Approximation (DWBA) using the code FOLD \cite{FOLD}.
FOLD is specifically designed to perform charge-exchange reaction calculations with composite probes: the form factor is created by double-folding the nucleon-nucleon interaction describing the interaction between nucleons in the target and projectile over the transition densities of the target-residual and projectile-ejectile systems. Shell-model calculations with the codes OXBASH \cite{OXBA} and NuShellX \cite{nushell} were used with appropriate effective interactions in the relevant model spaces to generate realistic sets of one-body transition densities based on modern shell-model interactions. Radial wave functions used in the form-factor calculations were typically generated in Wood-Saxon potentials, for which the well-depths were adjusted such that single-particle binding energies matched those calculated in the shell-model using the Skyrme SK20 interaction \cite{BRO98}. For the $t$ and $^{3}$He particles, radial densities obtained from Variational Monte-Carlo calculations \cite{WIR05} were used.

The effective nucleon-nucleon interaction of Love and Franey \cite{LOV81,LOV85} at 140 $A$MeV was used in the calculations of the form factors. The main deficiency in the cross-section calculations stems from the fact that a short-range approximation for the effects of the antisymmetrization of the di-nuclear system (the so-called `(knock-on) exchange' terms) must be used in the code FOLD. In Eq. (\ref{eq:dsigma}), the exchange contributions affect the value of $J_{\sigma \tau}$:
\begin{equation} \label{eq:je}
J_{\sigma \tau}=J^{D}_{\sigma \tau}+J^{E}_{\sigma \tau},
\end{equation}
where $D$ refers to the direct contribution and $E$ to the exchange contributions. An exact treatment of exchange effects for charge-exchange reactions with composite probes has only been performed for very specific cases \cite{UDA87,KIM00}, and a general tool to perform such calculations is not available.

Although the short-range approximation works reasonably well for nucleon-induced charge-exchange reaction calculations at intermediate energies, it is known to lead to an underestimation of exchange effects for reactions with composite probes \cite{UDA87,KIM00,HAG06}. Since the sign of the exchange contributions is opposite to the direct contributions, the underestimated exchange amplitudes give rise to a general overestimation of the calculated cross sections compared to the data \cite{ZEG07}.  Correcting for such effects is complicated since the exchange contributions are target-mass dependent \cite{LOV81,LOV85}. However, for transitions for which the $B(GT)$ is known from $\beta$-decay and the differential cross sections are extracted from experiment, the effect of the exchange contributions can be deduced if the parameters $K$ and $N^{D}$ in Eq. (\ref{eq:dsigma}) can be reliably calculated. Since the calculation of $K$ is trivial and the value of $N^{D}$ not very sensitive to the value of $|J_{\sigma \tau}|^{2}$, this is the case here and one of the goals of the present work is to establish a phenomenological description of $|J_{\sigma \tau}|$ as a function of mass number.

The $p$($t$,$^{3}$He)$n$ reaction discussed in Section \ref{sec:proton} is a special case. In the analysis of the data, DWBA calculations (for the inverse $t$($p$,$n$)$^{3}$He reaction) were performed with the code DW81 \cite{DW81}. This code is particularly well suited for nucleon-induced reactions, since exchange effects can be treated exactly instead of using the short-range approximation. However, in the analysis of the unit cross section for the $p$($t$,$^{3}$He)$n$ reaction in Section \ref{sec:jst}, calculations performed in DW81 and FOLD were both used and compared. In the case of the FOLD calculations, delta functions were used to describe the proton and neutron densities.

After a brief summary of the data used in this paper in section \ref{sec:Experimental}, new and reevaluated ($t$,$^{3}$He) data are discussed in more detail in section \ref{sec:newdata}. The generated results for the GT unit cross sections are then used to study the terms in the factorized expression of Eq. (\ref{eq:dsigma}) in Section \ref{sec:Results}.

\section{Data used in the analysis} \label{sec:Experimental}
The ($ ^{3} $He,t) data at 140 $A$MeV used in the current analysis are identical to those presented in Ref. \cite{ZEG07} and references therein. GT unit cross sections were extracted for transitions listed in Table \ref{tab:summary1}. The unit cross sections for nuclei with mass numbers $A$ of 62, 64 and 68 were derived from the relationship between Fermi and GT unit cross sections \cite{ADA06} and the empirical relationship between the unit cross section for Fermi transitions and mass number as described in Ref. \cite{ZEG07}. All these data were collected at the Research Center for Nuclear Physics in Osaka, using a beam of $^{3}$He$^{2+}$ particles at 140 $A$MeV. Tritons were analyzed in the Grand Raiden spectrometer \cite{FUJ99}.


Data from six (t,$ ^{3} $He) experiments performed at 115 $A$MeV were included in the current investigation. An overview of the transitions and their $B(GT)$ values is provided in the first 3 columns of Table \ref{tab:summary2}. All ($t$,$^{3}$He) experiments were performed at the Coupled Cyclotron Facility at NSCL. The $^{3}$He$^{2+}$ particles were analyzed in the S800 spectrometer \cite{BAZ03}. Results for experiments with $^{13}$C \cite{GUE09} and $^{26}$Mg \cite{ZEG06} targets have been published and we refer to the relevant publications for details. The analysis of the $^{6}$Li(t,$ ^{3} $He) reaction has also been published \cite{NAK00} but was reevaluated (see Section \ref{sec:newdata}). The new (t,$ ^{3} $He) data using $p$, $^{2}$H and $^{12}$C targets, is discussed in more detail in Section \ref{sec:newdata}.

\begin{table*}
\caption{\label{tab:summary1}Overview of extracted cross sections and unit cross sections for various transitions excited via the ($^{3}$He,$t$) reaction at 140 $A$MeV (see also Ref. \cite{ZEG07}. Indicated are the initial and final state, the $B(GT)$ associated with the transition, the extracted differential cross section at $0^{\circ}$ and the extrapolation to $q=0$,  the derived unit cross section and the reference. The $B(GT)$ values were calculated from known log$ft$ values \cite{ENSDF} following Eq. (\ref{eq:logft}), unless indicated otherwise.}
\begin{ruledtabular}
\begin{tabular}{llccccccc}
    $i$   & $f$ & $B(GT)$  & d$\sigma$/d$\Omega_{c.m.}$($0^{\circ}$) & d$\sigma$/d$\Omega_{c.m.}$($q=0$)  & $\hat{\sigma}$ & Ref. \\
          &            &    & (mb/sr) & (mb/sr) & (mb/sr) &  \\ \hline
    $^{12}$C($0^{+}$,g.s.) & $^{12}$N($1^{+}$,g.s.) &  0.88 &$16.1\pm0.12$ & $19.9\pm1.0$   & $22.6\pm1.1$ & \cite{ZEG07} \\
    $^{13}$C($1/2^{-}$,g.s.) & $^{13}$N($3/2^{-}$,15.1 MeV) &  $0.23\pm 0.01$ & $3.65\pm0.10$ & $4.51\pm0.26$ & $19.7\pm1.1$ & \cite{ZEG08} \\
    $^{18}$O($0^{+}$,g.s.) & $^{18}$F($1^{+}$,g.s.) & 3.11  & $51.2\pm2.2$ & $51.2\pm3.4$ &  $16.5\pm1.1$ & \cite{ZEG07} \\
    $^{26}$Mg($0^{+}$,g.s.) & $^{26}$Al($1^{+}$,1.06 MeV) & 1.1 & $13.9\pm0.3$ & $14.1\pm0.8$    & $12.8\pm0.7$ & \cite{ZEG06} \\
    $^{58}$Ni($0^{+}$,g.s.) & $^{58}$Cu($1^{+}$,g.s.) & 0.155 & $1.5\pm0.01$ & $1.5\pm0.08$  & $9.65\pm0.48\footnotemark[1]$ & \cite{ZEG07} \\
    $^{62}$Ni($0^{+}$,g.s.) & $^{62}$Cu($1^{+}$,g.s.) & 0.073& -     & -      & $7.7\pm1.0$\footnotemark[2] & \cite{ADA06,ZEG07} \\
    $^{64}$Ni($0^{+}$,g.s.) & $^{64}$Cu($1^{+}$,g.s.) & 0.123& -     & -      & $7.4\pm0.9$\footnotemark[2] & \cite{ADA06,ZEG07} \\
    $^{68}$Zn($0^{+}$,g.s.) & $^{68}$Ga($1^{+}$,g.s.) & 0.073& -     & -      & $7.0\pm0.8$\footnotemark[2] & \cite{ADA06,ZEG07} \\
    $^{118}$Sn($0^{+}$,g.s.) & $^{118}$Sb($1^{+}$,g.s.) & 0.344& $1.71\pm0.04$ & $1.62\pm0.09$  & $4.72\pm0.26$ & \cite{ZEG07} \\
    $^{120}$Sn($0^{+}$,g.s.) & $^{120}$Sb($1^{+}$,g.s.) & 0.345& $1.80\pm0.10$ & $1.72\pm0.13$  & $5.00\pm0.37$ & \cite{ZEG07} \\
          &       &       &       &       &       &  \\
\end{tabular}
\end{ruledtabular}
\footnotetext[1]{This transition is known to be strongly affected by interference between $\Delta L=0$ and $\Delta L=2$ amplitudes, see Section \ref{sec:Theory}. The effect was estimated to be 20\% \cite{COL06}, which reduces the unit cross section to $8.0\pm0.5$ mb/sr.}
\footnotetext[2]{Unit cross section established using the $R^{2}$ value from Ref. \cite{ADA06} and multiplying with ${\hat{\sigma}}_{F}$ from Ref. \cite{ZEG07}.}
\end{table*}

\begin{table*}
\caption{\label{tab:summary2}Overview of extracted cross sections and unit cross sections for various transitions excited via the ($t$,$^{3}$He) reaction at 115 $A$MeV. Indicated are the initial and final state, the $B(GT)$ associated with the transition, the extracted differential cross section at $0^{\circ}$ and the extrapolation to $q=0$,  the derived unit cross section and the reference. The $B(GT)$ values were calculated from known log$ft$ values \cite{ENSDF} following Eq. (\ref{eq:logft}), unless indicated otherwise.}
\begin{ruledtabular}
\begin{tabular}{llccccccc}
    $i$   & $f$   & $B(GT)$& d$\sigma$/d$\Omega$($0^{\circ}$) & d$\sigma$/d$\Omega$($q=0$)  & $\hat{\sigma}$ & Ref. \\
          &       & (mb/sr) & (mb/sr) &       & (mb/sr) &  \\ \hline
    $^{1}$H($1/2^{+}$) & $^{1}$n($1/2^{+}$) & 3     & $25\pm2$ & $25\pm2$ & $8.3\pm0.7$ & this work \\
    $^{2}$H($1^{+}$) & 2$n$($0^{+}$) & $-$  & $-$  & $-$  & $13.0\pm1.3$\footnotemark[1] & this work \\
    $^{6}$Li($1^{+}$, g.s.) & $^{6}$He($0^{+}$, g.s.) & 1.577& $51\pm4$ & $52\pm4$  & $32.9\pm2.6$ & \cite{NAK00}, reevaluated \\
    $^{12}$C($0^{+}$, g.s.) & $^{12}$B($1^{+}$, g.s) & 0.99& $16.6\pm1.2$ & $20.4\pm1.5$   & $20.5\pm1.5$ & this work \\
    $^{13}$C($1/2^{-}$, g.s.) & $^{13}$B($3/2^{-}$, g.s.) & 0.711& $13.1\pm1.3$ & $16.2\pm1.6$  & $22.8\pm2.3$ & \cite{GUE09} \\
    $^{26}$Mg($0^{+}$, g.s.) & $^{26}$Mg($1^{+}$, 0.08 MeV) & $0.41\pm0.02$\footnotemark[2]& $4.1\pm0.3$ & $5.27\pm0.4$  & $12.8\pm1.0$ & \cite{ZEG06} \\
\end{tabular}
\end{ruledtabular}
\footnotetext[1]{See Section \ref{sec:he}.}
\footnotetext[2]{Derived from combining with $^{26}$Mg($^{3}$He,$t$) data and applying isospin symmetry \cite{ZEG06}.}
\end{table*}



\section{New and reevaluated ($\textbf{t}$,$^{3}\textrm{\textbf{He}}$) data} \label{sec:newdata}

\subsection{The $p$($t$,$ ^{3} $He)$n$ reaction} \label{sec:proton}
\begin{figure}
\includegraphics[scale=1]{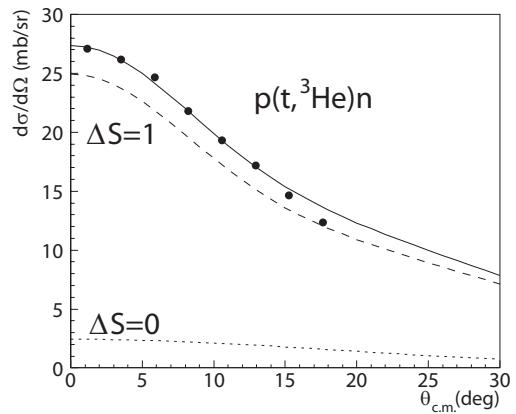}
\caption{\label{HydrogenExpTheory} Differential cross section for the $p$(t,$ ^{3} $He)$n$ reaction $E(t)=115$ $A$MeV. The theoretical angular distribution is fitted to the data, with the normalization being the only fit parameters. The relative contributions from transitions associated with $\Delta S=0$ and $\Delta S=1$ is fixed (see text).}
\end{figure}

Data on the $p$($t$,$ ^{3} $He)$n$ reaction was extracted using a 99.3\% isotopically-enriched $^{13}$CH$_{2}$ target with a thickness of 18.0 mg/cm$^{2}$. Events associated with the ($t$,$ ^{3} $He) reaction on hydrogen and $^{13}$C present in the target can be separated owing to the difference in ground-state $Q$-value of 12.6 MeV. The analysis of the $^{13}$C($t$,$^{3}$He) data has been discussed in Ref. \cite{GUE09} and we refer to that publication for the experimental details. In the same experiment, data was also taken with a $^{nat}$CH$_{2}$ target (see also section \ref{sec:12c}). Cross sections for the $p$(t,$ ^{3} $He)$n$ extracted from the two targets were consistent, but statistical errors were smaller with the $^{13}$CH$_{2}$ target. Therefore, this data set was used in the present analysis. The extracted differential cross section for the $p$($t$,$ ^{3} $He)$n$ reaction is shown in Fig. \ref{HydrogenExpTheory}.

A complication in the extraction of the cross section associated with the GT transition is that the $p$($t$,$ ^{3} $He)$n$ reaction has mixed Fermi ($ \Delta $S=0) and GT ($ \Delta $S=1) character. The transition from the proton to the neutron exhausts the full Fermi ($ B(F)=|N-Z|=1 $) and GT ($B(GT)=3|N-Z|=3 $) sum rules. Since both types of transitions are associated with angular momentum transfer $\Delta L=0$, the angular distributions of the differential cross section for the two types are nearly identical and the experimental results presented in Fig. \ref{HydrogenExpTheory} cannot easily be decomposed. However, the ratio of GT and Fermi unit cross sections $R^{2}=\frac{{\hat{\sigma}}_{GT}}{{\hat{\sigma}}_{F}}$ is accurately described for ($p$,$n$) charge-exchange reactions as a function of beam energy $E_{b}$,  for $E_{b}\lesssim 200$ MeV \cite{TAD87}:
\begin{equation}
\label{eq:R2}
R^{2}_{(p,n)}=\left[\frac{E_{b}}{E_{0}}\right]^{2}.
\end{equation}
The constant $E_{0}$ was empirically established to be $55.0 \pm 0.4$ MeV. At $E_{b}=115$ MeV, $R^{2}_{(p,n)}=4.37\pm0.06$. Therefore, the ratio of contributions from the GT and Fermi components to the experimental differential cross section shown in Fig. \ref{HydrogenExpTheory} is expected to be $R^{2}_{(p,n)}\frac{B(GT)}{B(F)}=13.1\pm0.2$. This ratio was used to fix the relative contributions from differential cross sections calculated for each of the components in DWBA. The summed theoretical angular distribution was then fitted to the data, with the absolute normalization as the only fit parameter. As mentioned, for the purpose of the $p$($t$,$ ^{3} $He)$n$ DWBA calculations, the code \textsc{DW81} \cite{DW81} was used. Optical potential parameters deduced from $p$+$t$ and $p$+$^{3}$He elastic scattering at 156.5 $A$MeV \cite{OER82} were used. Single-particle states in $^{3}$He and $^{3}$H were generated in a harmonic oscillator potential with oscillator parameter $b=1.4$ fm, following Ref. \cite{CEL93}. The effective interaction used in the calculations was the Love-Franey interaction \cite{LOV81,LOV85} at 140 MeV. All nucleons in the ground states of the $A=3$ nuclei were assumed to be in the $0s_{1/2}$ shell.

Taking into account errors in the beam normalization of 5\%, the total differential cross section for the $p$($t$,$ ^{3} $He)$n$ reaction at $0^{\circ}$ was found to be $27.0\pm 1.4$ mb/sr. Using the above-mentioned ratio for the contributions from the GT and Fermi transitions, the GT cross section at $0^{\circ}$ is $25\pm1.4$ mb/sr. Because the value of $E_{0}$ was established for ($p$,$n$) reactions with targets of mass number $A\geq7$ and the uncertainties for $A=3$ unknown, a systematic error was assigned to the extracted GT cross section with the conservative assumption that the Fermi contribution at $0^{\circ}$ could be off by as much as 50\% of the estimated value of 2 mb/sr. Therefore, the GT cross section at $0^{\circ}$ used in the further analysis was $25\pm2$ mb/sr (see Fig. \ref{HydrogenExpTheory}), from which a unit cross section for the ($t$,$ ^{3} $He) reaction on the proton of 8.3 $ \pm $ 0.7 mb/sr was deduced.

\subsection{The $d$($t$,$^{3}$He)2$n$ reaction}
\label{sec:deuteron}
The $d$($t$,$^{3}$He) data were taken during an experiment described in Ref. \cite{HIT09}, which focused on the study of the $^{64}$Zn($t$,$^{3}$He) reaction and we refer to the corresponding paper for the details of the measurement. The target was a deuterated polyethylene foil (CD$_2$) with a thickness of 9.1 mg/cm$^{2}$. The $^{12}$C present in the target was useful to cross-calibrate the absolute beam intensities between this earlier data set and the one described in Sections \ref{sec:proton} and \ref{sec:12c} by using the data from the $^{12}$C($t$,$^{3}$He) reaction. Whereas in the experiment described in Ref. \cite{HIT09} systematic uncertainties in the measurement of the absolute beam intensities were large ($\sim 40\%$), in the more recent measurement problems related to the beam integration were resolved and a much reduced uncertainty of $\pm5$\% was achieved. The scaling factor needed to normalize the older $^{12}$C($t$,$^{3}$He) data with the newer ones, was also applied in the analysis of the $d$($t$,$^{3}$He) reaction.

The ($t$,$^{3}$He) reaction on the deuteron populates the unbound $n+n$ system. Therefore, information about the GT transition strength cannot be obtained from $\beta$-decay data. However, the $B(GT)$ distribution for the $d\rightarrow n+n$ transition can be deduced from the $B$($M1$) distribution for the analog $d\rightarrow n+p$ transition. For the $d\rightarrow n+p$ transition, experimental information is available from the study of $\gamma d \rightarrow np$ \cite{SHI49,BIS50,BAR52,MOR89,SCH00,HAR03,TOR03}, $np \rightarrow d\gamma$ \cite{NAG97} and $d(e,e')$ \cite{RYE08} reactions. The experimental results agree well with calculations using effective field theories (EFT). The tabulated values from pionless EFT (EFT($\acute{\pi}$)) provided in Ref. \cite{CHE99} were used for the present analysis. $(n,p)$-type charge-exchange reactions on the deuteron have been studied in the past: Nakayama \textit{et al.} \cite{NAK05} used the ($^{7}$Li,$^{7}$Be) reaction and B\"{a}umer \textit{et al.} \cite{BAU05a} employed the ($d$,$^2$He) reaction. Here we follow a procedure quite similar to that of Ref. \cite{NAK05}, which is briefly summarized below.

It is assumed that the $J^{\pi}=1^{+}$ deuteron ground state is in a pure triplet $^{3}S$ ($T=0$) configuration (thus neglecting the 2.5\% $D$-state component \cite{ROD90}) and the ($t$,$^{3}$He) reaction on the deuteron populates the $^{1}$S, $T=1$, $J^{\pi}=0^{+}$, $n+n$ unbound state. Under these assumptions, the orbital contribution to the $M1$ transition can be neglected and $d\rightarrow n+n$ is the analog of the $d\rightarrow p+n$ transition. In this approximation, the $B(M1)$ for the $\gamma d\rightarrow np$ reaction is a direct measure for the GT transition strength associated with the isovector transition. It follows that:
\begin{equation}
\label{eq:db}
\frac{dB(M1)}{dE}=\frac{3(\mu_{p}-\mu_{n})^{2}}{8\pi}\frac{dB(GT)}{dE},
\end{equation}
where $\mu_p$ and $\mu_n$ are the magnetic moments of the proton and the neutron, respectively. The cross section for the $M1$ $\gamma$ transition ($\sigma_{M1}$) for $\gamma d\rightarrow n+p$ can be expressed in terms of $B(M1$) \cite{BOH,NAK05}:
\begin{equation}
\label{eq:m1}
\frac{d\sigma_{M1}}{dE_{\gamma}}(1^{+}\rightarrow 0^{+})=0.044 E_{\gamma} \frac{dB(M1)}{dE_{\gamma}}.
\end{equation}
Combining Eqs. \ref{eq:dsigma}, \ref{eq:db} and \ref{eq:m1} then provides the relation between $\sigma_{M1}$ and the cross section for the $d$($t$,$^{3}$He) charge-exchange reaction at $q=0$:
\begin{equation}
\label{eq:sm1}
\frac{d\sigma_{M1}}{dE_{\gamma}}(1^{+}\rightarrow 0^{+})=\frac{0.116E_{\gamma}}{\hat{\sigma}}\frac{d^{2}\sigma(q=0)}{dq dE}\Bigg\vert_{d(t,^{3}\textrm{He})}
\end{equation}
Therefore, given $\sigma_{M1}$, one can deduce $\hat{\sigma}$ from the experimental cross section for the $d$($t$,$^{3}$He) reaction.

\begin{figure}
\includegraphics[scale=1.0]{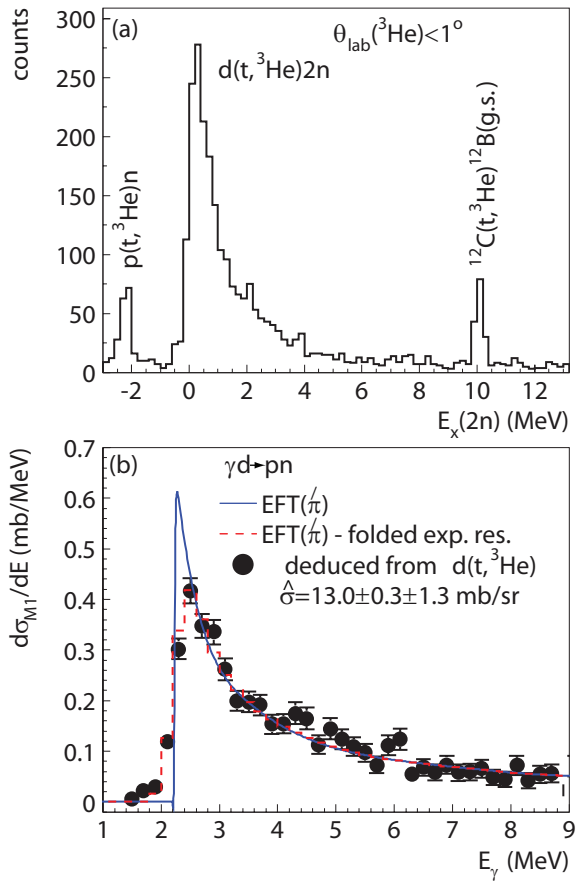}
\caption{\label{deuteron} (color online) (a) Excitation energy spectrum for the $d$($t$,$^{3}$He)$2n$ reaction for events with $\theta_{lab}(^{3}$He$<1^{\circ}$). Also visible are peaks due to the $^{12}$C($t$,$^{3}$He)$^{12}$B(g.s.) transition and the $^{1}$H($t$,$^{3}$He)n reaction. (b) Differential cross section for the $\gamma d\rightarrow p+n$ reaction based on EFT($\acute{\pi}$) (solid blue line), EFT($\acute{\pi}$) after folding with the energy resolution achieved in the $d$($t$,$^{3}$He)$2n$ experiment (dashed red line) and the measurement of $d$($t$,$^{3}$He)$2n$. The GT unit cross section $\hat{\sigma}$ for the ($t$,$^{3}$He) reaction on the deuteron was determined by fitting the experimental distribution to the theory.}
\end{figure}

The measured excitation-energy spectrum for the $d$($t$,$^{3}$He) reaction is shown in Fig. \ref{deuteron}(a), gated on events with a $^{3}$He laboratory scattering angle of less than $1^{\circ}$. Because the $n+n$ system is unbound, a broad energy distribution for the $n+n$ system is found. A peak due to the $^{12}$C($t$,$^{3}$He)$^{12}$B(g.s.) transition can be seen at about 10 MeV. In addition, a minor remaining amount of $^{1}$H in the CD$_2$ target caused a small peak at energies below that for the $d$($t$,$^{3}$He) distribution due to $^{1}$H($t$,$^{3}$He)$n$ reactions. The resolution in $E_{x}(2n)$ depends on: (i) the difference in energy loss of the $t$ and $^{3}$He in the target, (ii) the intrinsic resolution of the energy measurement and (iii) the resolution in the $^{3}$He scattering angle (which correlates with the recoil energy of the di-neutron system). The first contribution can be calculated and the second and third contributions were determined from the resolutions achieved for the $^{12}$C($t$,$^{3}$He)$^{12}$B(g.s.) transition also present in the data. For $\theta_{lab}(^{3}$He$<1^{\circ})$, the resolution in $E_{x}(2n)$ was 400 keV (FWHM).

Before Eq. (\ref{eq:sm1}) can be applied to deduce $\sigma_{M1}$ from the measured $d$($t$,$^{3}$He) cross sections, the data have to be extrapolated to $q=0$. The linear momentum transfer $q$ increases with $E_{x}(2n)$ and scattering angle. The multiplicative extrapolation factors were estimated in DWBA using the code FOLD \cite{FOLD}. They ranged from 1.05 for $E_{x}(2n)=0$ MeV to 1.15 at $E_{x}(2n)=9$ MeV. Since no empirical optical potentials are available for the $t+d$ and $^{3}$He+$2n$ channels, calculations with a variety of optical potentials were performed: a plane-wave calculation, calculations using optical parameters for $t+p$ (see section \ref{sec:proton}) and ($t$+$^{6}$Li) (see section \ref{sec:he}) and various interpolations of these potentials. Although the choice of optical potential strongly affects the absolute calculated cross sections, the extrapolation factor from finite $q$ to $q=0$ changed by at most 2\%. In these calculations, the radial wave function for the deuteron was based on the parametrization given in Ref. \cite{KRU07}. For the purpose of the DWBA calculation, the neutrons in the di-neutron system were initially assumed to be in a bound state with radial wave functions equal to that of the deuteron. The sensitivity of the extrapolation from finite $q$ to $q=0$ on the choice these wave functions was tested by repeating the calculations with wave functions extending to larger radii as expected for the unbound system. The effect on the extrapolation factors was $\sim 1$\%. We concluded that although the uncertainties in the inputs for the DWBA calculations are quite large, the extrapolation factors to $q=0$ have small errors.

In Fig. \ref{deuteron}(b), $\sigma_{M1}$ calculated in EFT($\acute{\pi}$) is plotted as a function of $E_{\gamma}$ for the $\gamma d \rightarrow pn$ reaction (solid blue line). The threshold for this reaction is 2.24 MeV (the binding energy of the deuteron). The EFT($\acute{\pi}$) curve was folded with the experimental energy resolution (dashed red line). To determine the unit cross section $\hat{\sigma}$ in Eq. (\ref{eq:sm1}), the $\sigma_{M1}$ distribution deduced from the $d$($t$,$^{3}$He) data was then fitted to the EFT($\acute{\pi}$) curve that was folded with the experimental energy resolution. $\hat{\sigma}$ was the only fit parameter. A value of $13.0\pm0.3$ mb/sr was found ($\chi^{2}/n=1.16$ with $n=37$). Based on the assumptions made about the reaction mechanism and the uncertainties in various experimental parameters and the EFT($\acute{\pi}$) calculations, we estimated that the systematic error was about 10\% of this value.

\subsection{The $ ^{6} $Li(t,$ ^{3} $He)$ ^{6} $He(g.s.) reaction}
\label{sec:he}
\begin{figure}
\includegraphics[scale=1.0]{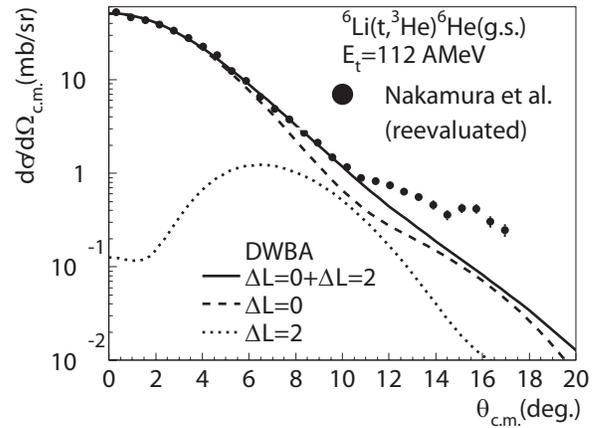}
\caption{\label{6he} Differential cross section for the $^{6}$Li($t$,$^{3}$He(g.s.) reaction at 112 AMeV. The experimental data have been reevaluated as discussed in the text. The full solid lines shows the result of the DWBA calculation; the contributions from the $\Delta L=0$ (dashed line) and $\Delta L=2$ component (dotted line) were deduced in a fit.}
\end{figure}

The $^{6}$Li($1^{+}$,g.s.)$\rightarrow^{6}$He($0^{+}$,g.s.) transition has a known $B(GT)$ of 1.577 \cite{TIL02} from $\beta$-decay data and can thus be used to extract a unit cross section. Because of concerns about systematic uncertainties in the beam normalization, the absolute scale of the differential cross section for this transition reported in Ref. \cite{NAK00} was set by comparing data taken for the $^{12}$C($t$,$^{3}$He)$^{12}$B(g.s.) reaction in the same experiment, with data from an earlier experiment for that reaction \cite{DAI97}. However, in the earlier experiment, cross sections were integrated over a relatively large solid angle and thus did not provide an accurate measure for the differential cross section near $0^{\circ}$. The differential cross section for the $^{12}$C($t$,$^{3}$He)$^{12}$B(g.s.) $0^{\circ}$ transition taken in the same experiment as the $^{6}$Li data was measured to be $15.4\pm0.9$ mb/sr \cite{NAK09}. Since the value reported in Ref. \cite{DAI97} was much lower ($11.8\pm1.4$ mb/sr), the measured differential cross sections for the $^{6}$Li($t$,$^{3}$He) were scaled down accordingly in Ref. \cite{NAK00}. However, the original number of $15.4\pm0.9$ mb/sr is within error margins consistent with our new result for the $^{12}$C($t$,$^{3}$He)$^{12}$B(g.s.) transition (see Section \ref{sec:12c}). We therefore decided to rescale the reported differential cross section for the  $^{6}$Li($t$,$^{3}$He)$^{6}$He(g.s.) reaction in Ref. \cite{NAK00} by using the $^{12}$C($t$,$^{3}$He)$^{12}$B(g.s.) transition as a reference. The result is shown in Fig. \ref{6he}.

The experimental differential cross sections were compared with DWBA calculations by using the code FOLD \cite{FOLD}. Following Ref. \cite{NAK00}, one-body transition densities (OBTDs) were calculated in OXBASH \cite{OXBA} using the CKHE interaction \cite{STE88} in the $p$-shell model space. The CKHE interaction is a modified version of the CKI interaction \cite{COH65} that reproduces the binding and excitation energies of the He isotopes \cite{STE88}. Radial wave functions were generated in OXBASH as well, by using the SK20 interaction \cite{BRO98} while forcing the binding energies of the $p_{3/2}$ protons(neutrons) in $^{6}$Li($^{6}$He) to match the experimental values. The optical potential parameters determined from elastic scattering of $^{3}$He on $^{6}$Li \cite{NAK00} were used for the in and outgoing scattering channels.

To determine the experimental cross section at $0^{\circ}$, the DWBA calculations were scaled to fit to the data, as shown in Fig. \ref{6he}. The relative contribution from $\Delta L=0$ (GT) and $\Delta L=2$ components were deduced in a fit with a linear combination of the two components. A good correspondence between data and theory was achieved up to $\theta_{c.m.}=11^{\circ}$; at higher angles the DWBA calculation underestimates the data. The $0^{\circ}$ cross section  extracted was $51\pm4$ mb/sr ($\Delta L=0$ component only), where the error also indicates the uncertainty in the absolute normalization. The DWBA calculation was used to extrapolate this value to the $q=0$ limit with a result of $52\pm4$ mb/sr. By dividing this number with the known $B(GT)$ of 1.577, a GT unit cross section of $32.9\pm2.5$ mb/sr was deduced.

\subsection{The $ ^{12} $C(t,$ ^{3} $He)$ ^{12} $B(g.s.) reaction}
\label{sec:12c}
\begin{figure}
\includegraphics[scale=1]{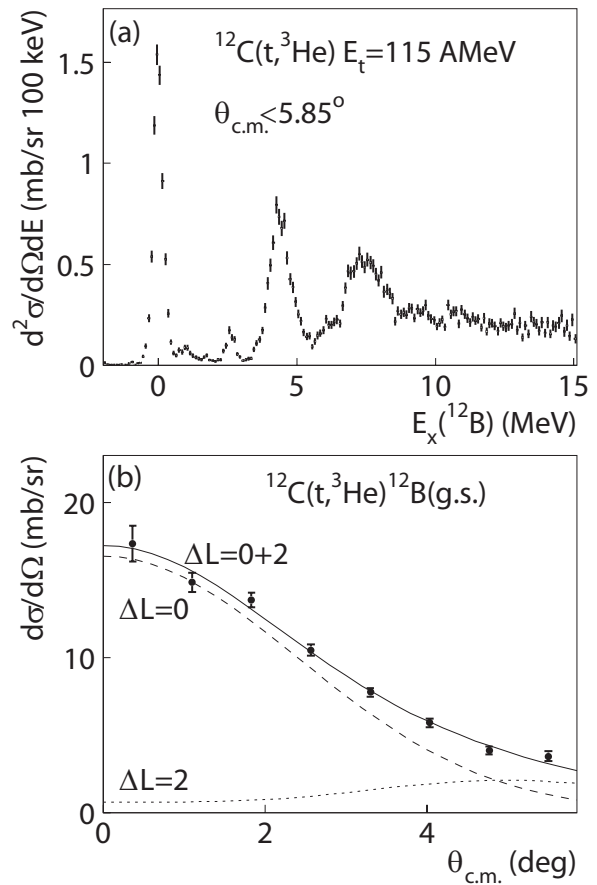}
\caption{\label{12c} (a) Excitation-energy spectrum for the $ ^{12}$C($t$,$^{3}$He) reaction at 115 $A$MeV. (b) Differential cross section for the $ ^{12} $C(t,$ ^{3} $He)$ ^{12} $B(g.s.) reaction. The data are compared with the DWBA calculation (solid line), in which the total cross section is decomposed in $\Delta L=0$ (dashed line) and $\Delta L=2$ (dotted line) contributions.}
\end{figure}

The $ ^{12} $C($t$,$ ^{3} $He) reaction was measured in the same experiment as the $ ^{13} $C($t$,$ ^{3} $He) reaction. The results of the latter reaction were published in Ref. \cite{GUE09} and we refer to that paper for the experimental details. A 10.0 mg/cm$ ^{2} $ thick  $^{nat}$CH$ _{2}$ target was used. The measured excitation-energy spectrum of $^{12}$B is shown in Fig. \ref{12c}(a). The prominent peak at 0 MeV corresponds to the $ ^{12} $C(0$ ^{+},g.s.) \rightarrow$ $ ^{12} $B(1$ ^{+}$,g.s.) transition, for which the $B(GT)=0.99$ is known from $\beta$-decay data. The extracted differential cross section for this transition is shown in Fig. \ref{12c}(b). The $\Delta L=0$ (GT) contribution was extracted by decomposing the measured differential cross section in $\Delta L=0$ and $\Delta L=2$ contributions. The code FOLD \cite{FOLD} was used to calculate the theoretical angular distributions. OBTDs were calculated in OXBASH \cite{OXBA} using the CKII interaction \cite{COH67} in the $p$-shell model space. Optical potential parameters were taken from Ref. \cite{KAM03}. Following Ref. \cite{WER89}, well-depths of the real and imaginary potentials for the $t$+$^{12}$C channel were set to 85\% of the well-depths for the $^{3}$He+$^{12}$B channel.

The result of the multipole decomposition is shown in Fig. \ref{12c}(b). The extracted cross section at $0^{\circ}$ for the $\Delta L=0$ component is 16.55$ \pm $1.2 mb/sr. The error includes a statistical component and a systematic component related to the absolute normalization of the cross section. Owing to the relatively large $Q$-value of -17.357 MeV for this transition, the effect of the extrapolation of this cross section to $q=0$ is significant. Using Eq. (\ref{eq:extrapolation}), a value of 20.4$ \pm $1.5 mb/sr is found, resulting in a unit cross section of 20.5$ \pm $1.5 mb/sr. This is close to the unit cross section extracted via the analog $^{12}$C($^{3}$He,$t$) reaction \cite{ZEG08}, for which a value of 22.6$\pm$1.1 mb/sr was found. It is another confirmation that the ($t$,$^{3}$He) reaction at 115 $A$MeV is very similar to the ($^{3}$He,$t$) reaction at 140 $A$MeV, in spite of the slight difference in beam energy.

\subsection{Summary of unit cross sections.}
\label{sec:summary}

Tables \ref{tab:summary1} and \ref{tab:summary2} provide overviews of extracted cross sections and unit cross sections for transitions studies via the ($^{3}$He,$t$) reaction at 140 $A$MeV and the ($t$,$^{3}$He) reaction at 115 $A$MeV. The tables give both the differential cross sections at $\theta_{c.m.}=0^{\circ}$ and the extrapolated values at $q=0$.  The value of the latter is divided by $B(GT)$ to determine the unit cross section. For the cases where the unit cross section is derived using a different method, the tables indicate the methods used.

\begin{figure}
\includegraphics[scale=1.]{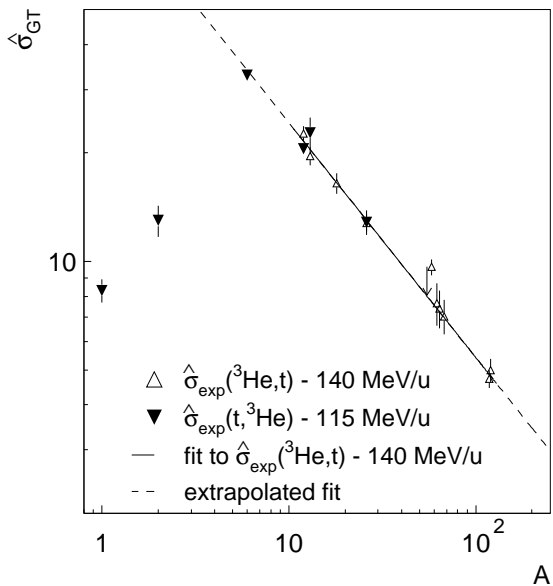}
\caption{\label{sigma_hat} Unit cross sections obtained from ($^{3}$He,$t$) and ($t$,$^{3}$He) experiments at 115 AMeV and 140 AMeV, respectively. The solid line show a fit to the ($^{3}$He,$t$) data for $12\leq A\leq 120$, as discussed in Ref. \cite{ZEG07}. The dashed line shows the extrapolation of the fitted function towards lower and higher mass numbers.}
\end{figure}

In Fig. \ref{sigma_hat}, the extracted unit cross sections are plotted as a function of mass number. The solid line indicates the fit to ($^{3}$He,$t$) unit cross sections \cite{ZEG07} for $A\geq12$:
\begin{equation} \label{eq:phen}
{\hat{\sigma}}_{GT}=\frac{109}{A^{0.65}}
\end{equation}
For reasons discussed in \ref{sec:Theory}, the unit cross section extracted from the $^{58}$Ni($^{3}$He,$t$) reaction is larger than the fitted curve. The arrow for this data point indicates the correction estimated based on theoretical calculations for the effect of the tensor interaction.
In the further analysis presented in this paper, this corrected value, which corresponds well with the fitted curve, is used.
The unit cross sections extracted from ($t$,$^{3}$He) experiments with target masses greater or equal to 12 are also consistent with Eq. (\ref{eq:phen}). The unit cross section for the $^{6}$Li($t$,$^{3}$He) reaction also lies on this function, even though $A=6$ is outside of the mass region considered in the fit. The unit cross sections for the ($t$,$^{3}$He) reactions on the proton and deuteron are much lower (by factors of 13 and 5, respectively) than the values expected based on Eq. (\ref{eq:phen}).

\section{Analysis of the unit cross sections and their dependence on target mass number}
\label{sec:Results}
In this section the phenomenological mass dependence of the unit cross section as a function of mass number shown in Fig. \ref{sigma_hat} is analyzed in terms of the factorization of Eq. (\ref{eq:dsigma}).

\subsection{Kinematic factor $K$}
\begin{figure}
\includegraphics[scale=1]{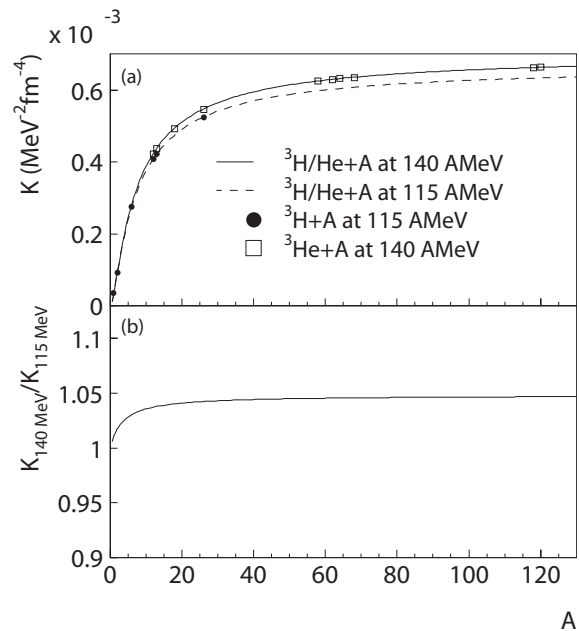}
\caption{\label{K_factor}(a) Dependence of the kinematic factor $K$ as defined in Eq. (\ref{eq:Kfactor}) for $t$ and $^{3}$He beam energies of 115 $A$MeV and 140 $A$MeV, respectively. The value of $K$ for target mass number used in the current work is indicated by markers. (b) Ratio of the the kinematic factors at beam energies of 140 $A$MeV and 115 $A$MeV.}
\end{figure}

The kinematic factor can be calculated analytically as a function of mass number using Eq. (\ref{eq:Kfactor}) and is shown in Fig. \ref{K_factor}(a). The markers correspond to the values of K for mass numbers studied experimentally in this work. A small difference in $K$ is present for the experiments performed at 115 $A$MeV and 140 $A$MeV. The magnitude of the difference is less than 5\% (see Fig. \ref{K_factor}(b)) which is smaller than typical uncertainties in the unit cross sections extracted from the data.  The kinematic factor rises rapidly with mass $A\lesssim40$.

\subsection{Distortion Factor $N^{D}$ } \label{subsec:Distortion}

The distortion factor $N^{D}$ is calculated following Eq. (\ref{eq:dist}). The PWBA and DWBA calculations are identical except that the depths of the optical potentials and the charges of the nuclei involved are set to zero in the PWBA calculations. Optical potential parameters are typically derived from fitting theoretical calculations using a fixed optical potential model to elastic scattering data. The fitting procedure is associated with statistical uncertainties and systematical errors due to uncertainties in the beam normalization and the target thickness. In addition, the choice of the terms used in the optical potential model can lead to systematic errors. Further uncertainties arise from the fact that elastic scattering data are not available for all nuclei studied, and optical potential parameters must be used from nuclei with similar mass numbers or by interpolating parameters from two or more target nuclei with similar mass numbers. An additional complication is that optical potential parameters are not available for the scattering of tritons on nuclei at beam energies exceeding 100 $A$MeV. For the $^{3}$He particles almost all elastic scattering data was taken at 140-150 $A$MeV \cite{YAM95,KAM03,FUJ04}.  Following Ref. \cite{WER89}, one usually adjusts the depths of the triton optical potentials to be 85\% of the ones for the $^{3}$He particles. For the ($t$,$^{3}$He) experiments performed at 115 $A$MeV, the parameters derived from elastic scattering at 140-150 $A$MeV were used. Given the consistency between unit cross sections for the ($t$,$^{3}$He) and ($^{3}$He,$t$) reactions, one can conclude that this procedure is reliable.

An estimate for the uncertainty in the distortion factors can be based on their functional form in the Eikonal approximation \cite{TAD87}:
\begin{equation} \label{eq:eikdist}
N^{D}=\exp(-xA^{1/3}+a_{0}),
\end{equation}
where
\begin{equation} \label{eq:eikdist1}
x=\frac{4Wr}{\hbar c \beta}.
\end{equation}
The parameters $r$ and $W$ are the radius and depth of the imaginary part of the optical potential, respectively, $\beta$ the velocity of the projectile and $a_{0}$ accounts for the difference between the depth, radius and velocity of the imaginary part of the optical potentials for the in and outgoing channels. Although the approximation of Eq. (\ref{eq:eikdist}) is too rough to accurately calculate distortion factors, it indicates that the uncertainty in their values is dominated by the product of the depth and the radius of the imaginary part of the optical potential. However, $Wr$ is rather stable as a function of mass number and beam energy for the ($^{3}$He,$t$) reaction, as can be seen by comparing Ref. \cite{KAM03} ($E(^{3}$He=148 $A$MeV)) and Ref. \cite{WIL73} ($E(^{3}$He=72 $A$MeV)). Moreover, these parameters cannot be changed drastically without degrading the overall good description of the angular distributions for excitations through the ($^{3}$He,$t$) and ($t$,$^{3}$He) data calculated in DWBA. On the basis of a sensitivity study, we estimated that the uncertainties in the calculation of the distortion factors are about 10\%.

\begin{figure}
\includegraphics[scale=1]{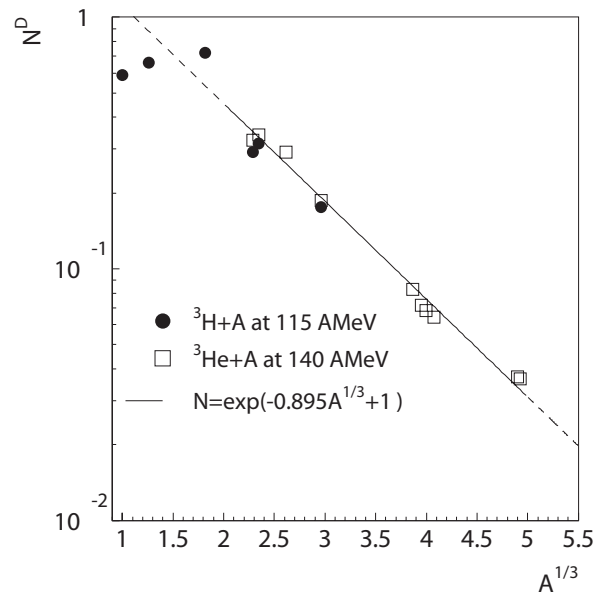}
\caption{\label{distortion}The distortion factor $N^{D}$ as defined in equation \ref{eq:dist} for ($^{3}$He,$t$) (solid circles)(open squares) and ($t$,$^{3}$He) (solid circles) reactions considered in this work. The solid line represents the result of a fit with an a function of the form expected for the distortion factor in the Eikonal approximation (see Eq. (\ref{eq:eikdist}) and text) for $A\geq 12$ The distortion factor for the $d$($t$,$^{3}$He)$2n$ carries a large systematic error of 0.15 (see text).}
\end{figure}

In Fig. \ref{distortion}, the distortion factors derived from Eq. (\ref{eq:dist}) are plotted as a function of $A^{1/3}$. For $A\geq12$, there is a strong correlation between $A^{1/3}$ and $N^{D}$; within the uncertainty of about 10\% the calculated values are consistent with a function of the form of Eq. (\ref{eq:eikdist}) with $x=0.895$ and $a_{0}=1.0$.
We note that the values of these parameters are 60-80\% higher than the corresponding parameters for the ($p$,$n$) reaction at similar beam energies \cite{TAD87}, yielding distortion factors for the ($t$,$^{3}$He) reaction about 20\% of those for the ($p$,$n$) reaction.

For the $d$($t$,$^{3}$He)$2n$ reaction measured optical potentials are not available (see Section \ref{sec:deuteron}) and the distortion factor could not be reliably calculated. The distortion factor for this reaction displayed in Fig. \ref{distortion} is an average of the values calculated with optical model parameters for the $^{3}$He+$p$ and $^{6}$Li+$^{3}$He reactions and has an uncertainty of 0.15 (23\%). The distortion factors for the ($t$,$^{3}$He) reactions on the proton and $^{6}$Li also deviate strongly from the trend line valid for the higher masses.

\subsection{Volume Integral of the interaction, $|J_{\sigma \tau}|$}
\label{sec:jst}
\begin{figure}
\includegraphics[scale=1.0]{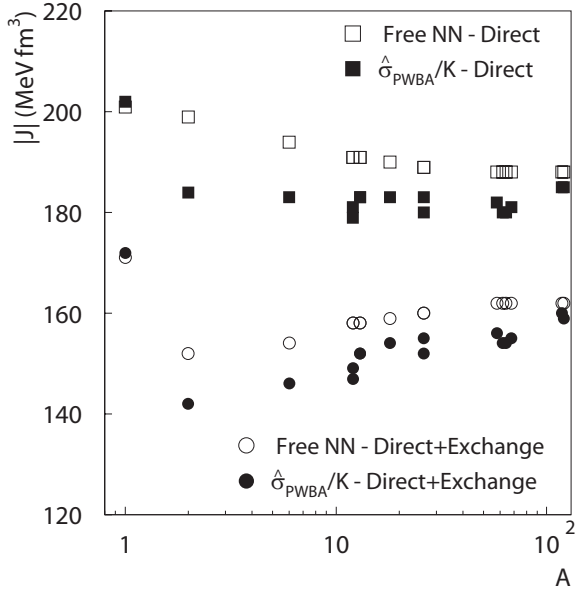}
\caption{\label{j} Calculations of the volume integral of the effective interaction using the Love-Franey parametrization \cite{LOV85}. Open squares represent a calculation using the free nucleon-nucleon interaction without any exchange effects taken into account. Open circles correspond to the calculation using the free nucleon-nucleon interaction including the zero-range approximation for the description of exchange effects. Solid squares (without exchange contribution) and circles (with exchange contributions) refer to the calculations in which the Love-Franey interaction was double folded over the transition densities of the target and projectile systems. Note the strongly suppressed zero on the ordinate.}
\end{figure}

The volume integral of the effective $\sigma\tau$ interaction responsible is the third parameter of importance for calculating the GT unit cross section. As mentioned in section \ref{sec:Theory}, a parametrization of the free nucleon-nucleon $t$-matrix by Love and Franey \cite{LOV81,LOV85} can be used to calculate $|J_{\sigma \tau}|$. The $t$-matrix is conveniently tabulated so that linear combinations of its parameters can be applied directly to scattering processes associated with the transfer of definite quanta of spin and isospin. The free nucleon-nucleon interaction of Refs. \cite{LOV81,LOV85} must be transformed from the nucleon-nucleon system to the nucleon-nucleus or nucleus-nucleus system (such as the ($t$,$^{3}$He) and ($^{3}$He,$t$) reaction discussed in this work) \cite{LOV81,KHO04}, which results in a renormalization of $|J_{\sigma \tau}|$ with a weak target-mass dependence, even if only direct terms of the interaction are considered. This dependence is shown in Fig. \ref{j} by open square symbols. It causes $|J_{\sigma \tau}|$ to drop from 200 MeVfm$^{3}$ for $A=1$ to $\sim190$ MeVfm$^{3}$ for $A>100$.

For reactions with composite probes, the Love-Franey interaction must be double-folded over the transition densities of the nuclei involved in the reaction, thereby effectively changing the ranges of the different components of the interaction and resulting in a modification of $|J_{\sigma \tau}|$. The effect of the double-folding of the interaction was estimated in PWBA calculated with the code FOLD, by using a modified version of Eq. (\ref{eq:sigma_hat}):
\begin{equation}\label{eq:freefold}
\hat{\sigma}_{GT,\textrm{PWBA}} =K|J_{\sigma \tau}|^{2}.
\end{equation}
In this equation, the left-hand side corresponds to the calculated cross section in PWBA. $K$ and $J_{\sigma \tau}$ are the same as in Eq. (\ref{eq:sigma_hat}) and $N^{D}$ is set to unity since distortions are absent in the plane-wave calculation. Eq. (\ref{eq:freefold}) was used to solve for $|J_{\sigma \tau}|$. The results of this calculation are indicated by filled squares in Fig. \ref{j}. Exchange contributions are neglected at this point. Except for $A=1$ (the double folding by definition has no effect in that case), the folding of the interaction over the densities leads to an overall reduction of $|J_{\sigma \tau}|$ and a flattening of the target-mass dependence compared to the calculation using the free nucleon-nucleon interaction.

As discussed in section \ref{sec:Theory}, taking into account the effects of exchange contributions is complex for composite probes. The use of the short-range approximation presented in Refs. \cite{LOV81,LOV85} works reasonably well for nucleon-nucleus scattering processes, but is known to result in overestimates of the scattering cross sections for ($^{3}$He,$t$) reactions on nuclei \cite{UDA87}. To obtains a systematic picture of this effect, we compared calculated values of $|J_{\sigma \tau}|$ using the short-range approximation for the exchange terms with those extracted from the available ($^{3}$He,$t$) and ($t$,$^{3}$He) data. We used the formalism described in Ref. \cite{LOV81}, except that the calculation of $k_{A}$ (originally the momentum of the incident nucleon in the nucleon-nucleus system), was modified to account for the transformation to the nucleus-nucleus system.

Values of $|J_{\sigma \tau}|$ calculated from the free nucleon-nucleon interaction of Refs. \cite{LOV81,LOV85} are indicated in Fig. \ref{j} with open circles and should be compared to points indicated with the open squares that do not include the exchange contributions. The exchange amplitudes interfere destructively with the direct amplitudes and $|J_{\sigma \tau}|$ decreases significantly. The reduction of $|J_{\sigma \tau}|$ is stronger for light target nuclei.
Taking into account exchange contributions combined with the double folding of the interaction over the transition densities results in a further reduction of $|J_{\sigma \tau}|$, as indicated by filled circular markers in Fig. \ref{j}. This final set of calculations can be directly compared to values of $|J_{\sigma \tau}|$ extracted from experiment.

\begin{figure}
\includegraphics[scale=1.0]{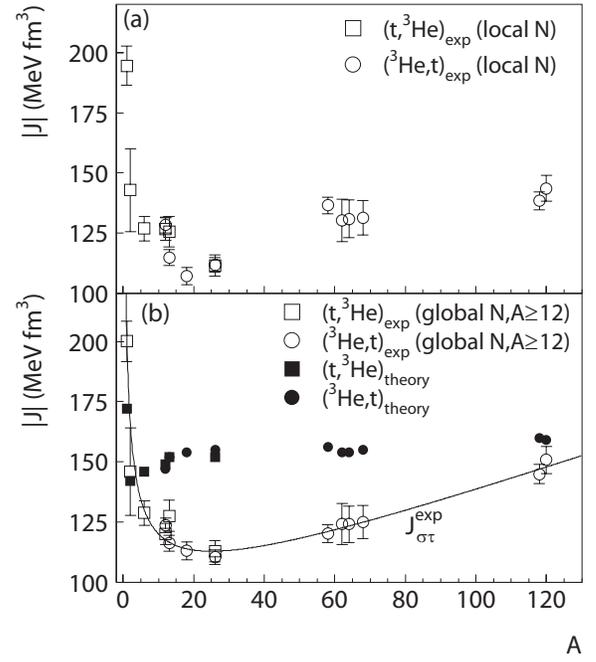}
\caption{\label{dj} (a) Extracted volume integrals of the effective interaction for the ($^{3}$He,$t$) (open circles) and ($t$,$^{3}$He) (open squares) reactions from the data by using Eq. (\ref{eq:sigma_hat}) and experimental GT strengths and differential cross section extrapolated to $q=0$. Local distortion factors are used. Note the strongly suppressed zero on the ordinate. (b) Idem, but with global distortion factors for $A\geq 12$. The solid line is a fit (Eq. (\ref{eq:jexp})) to these extracted values. In addition, the theoretical values for $|J_{\sigma\tau}|$ are shown for comparison.}
\end{figure}

Given the unit cross sections extracted from the ($^{3}$He,$t$) and ($t$,$^{3}$He) data, the calculated distortion factors $N^{D}$ and kinematical factors $K$,  $|J_{\sigma\tau}|$ can be deduced by using Eq. (\ref{eq:sigma_hat}). The results are shown in Fig. \ref{dj}(a). The distortion factors used are `local', i.e. calculated for each reaction separately, rather then using the trend line for $A\geq12$ shown in Fig. \ref{distortion}. The extracted values of $|J_{\sigma\tau}|$ vary from 105-140 MeVfm$^{3}$, except for the ($t$,$^{3}$He) reaction on the proton ($|J_{\sigma\tau}|=195$ MeVfm$^{3}$). A minimum is found near $A=20$.

Under the assumption that some scatter of the extracted values is caused by the uncertainties in the distortion factors, $|J_{\sigma\tau}|$ was recalculated by replacing local distortion factors for reactions involving targets with $A\geq12$ with the mass-dependent trend line shown in Fig. \ref{distortion}. The results (referred to as `global') are displayed in Fig. \ref{dj}(b). Distortion factors for the ($t$,$^{3}$He) reactions on the proton, deuteron and $^{6}$Li were left unchanged from their local values.
As a result of using the trend line for the distortion factors, the dependence of $|J_{\sigma\tau}|$ on $A$ smoothes and is well reproduced with the following purely phenomenological fit function, which is also included in Fig. \ref{dj}(b):
\begin{equation} \label{eq:jexp}
|J_{\sigma\tau}^{\textrm{exp}}|=\frac{128.5}{\sqrt{A}}+0.515A+74.3 \textrm{ for }A\leq120
\end{equation}

Also shown in Fig. \ref{dj}(b) are the calculated values of $|J_{\sigma\tau}|$, taking into account direct and exchange contributions and the double folding of the interaction over the transition densities of the projectile and target nuclei (i.e. the values indicated with solid circular markers in Fig. \ref{j}). Except for $A=1$, the calculated values are 15-30\% larger than the values extracted from the data and the discrepancy is largest for $A\approx20$. There is some qualitative consistency between the theoretical and experimental dependencies: for both, a rapid decrease of $|J_{\sigma\tau}|$ with increasing mass number is seen for low mass numbers, followed by a gradual increase for larger mass numbers. However, the minimum value of $|J_{\sigma\tau}|$ is reached near $A=2-6$ in the theoretical calculations, whereas it appears at $A\sim26$ in the results deduced from the experiments.

The experimentally extracted value of $|J_{\sigma\tau}|$ for the $p$($t$,$^{3}$He) reaction is higher than the theoretical estimate. In fact, it is close to the theoretical calculation in which exchange contributions are neglected. To check the theoretical estimate for this reaction, calculations were also performed using the code DW81, which allows for the exact treatment of exchange contributions, rather than the short-range approximation applied in FOLD. Results from DW81 for the distortion factor and $|J_{\sigma\tau}|$ were found to be nearly identical (deviations of less than 3\% were found, presumably because the single-particle wave functions of the nucleons in the triton and $^{3}$He were generated in a harmonic oscillator potential, rather than using the results from Variational Monte Carlo calculations). The close correspondence confirms the appropriateness of the short-range approximation of the exchange contribution for nucleon-nucleus scattering (i.e. for ($p$,$n$) reactions) but does not explain the anomalously high value of $|J_{\sigma\tau}|$ found for the $p$($t$,$^{3}$He)$n$ reaction. It is not inconceivable that the distortion factor has a larger error than estimated, especially since the optical model parameters for the $p$+$^{3}$He channel vary rapidly and non-uniformly as a function of beam energy \cite{OER82}. If the distortion factor is larger than calculated using Eq. (\ref{eq:dist}) by about 15\%, the extracted value of $|J_{\sigma\tau}|$ from the data would decrease to the value predicted taking into account the exchange contributions.

\begin{figure}
\includegraphics[scale =1]{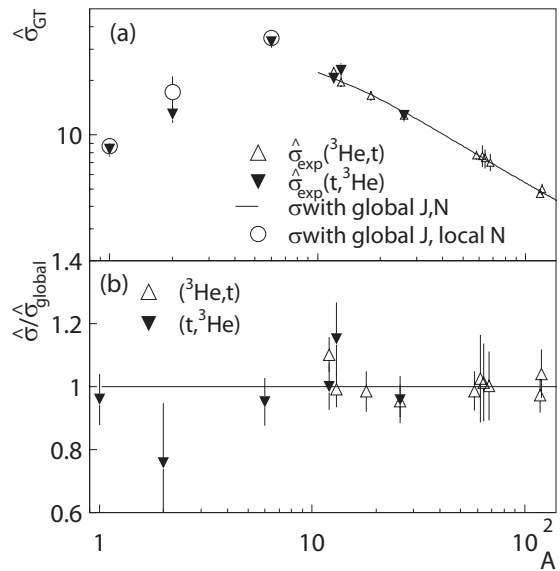}
\caption{\label{prop2} (a)Experimentally extracted unit cross sections for the ($^{3}$He,$t$) (upright open triangles) and ($t$,$^{3}$He) (inverted filled triangles) reactions are compared with the calculated unit cross sections using the (phenomenological) functions for $K$, $N^{D}$ and $|J_{\sigma\tau}|$. For $A<12$, local distortion factors are used instead of equation Eq. (\ref{eq:eikdist}. (b) The ratio of unit cross sections extracted from the data to the unit cross sections calculated with the (phenomenological) functions for $K$, $N^{D}$ and $|J_{\sigma\tau}|$.}
\end{figure}

\subsection{Synopsis}
\label{sec:overview}
With the phenomenological description of $|J_{\sigma\tau}|$ given in Eq. (\ref{eq:jexp}), the simple functional form of the distortion factor in Eq. (\ref{eq:eikdist}) with $x=0.895$ and $a_{0}=1$ and the kinematical factor $K$ in Eq. (\ref{eq:Kfactor}), one can calculate the GT unit cross section for the ($t$,$^{3}$He) and ($^{3}$He,$t$) reactions on targets with $A\geq12$. The results of this calculation are compared with the experimentally extracted unit cross sections in Fig. \ref{prop2}(a). For targets with $A<12$, the distortion factor was calculated separately, but the phenomenological description of $|J_{\sigma\tau}|$ can still be applied. In Fig. \ref{prop2}(b), the ratio of the experimental GT unit cross sections and the phenomenological description is shown, indicating a typical deviation between the two of less than 10\%. As explained in section \ref{sec:jst}, the good correspondence for $A=1$ could be due to the fact that an underestimate of the distortion factor $N^{D}$ led to an overestimate of $|J_{\sigma\tau}|$. The experimentally extracted unit cross section for $A=2$ (deuteron) also has a large systematic error due to the difficulty in calculating the distortion factor. Finally, we note that the experimental unit cross section for $A=6$ ($^{6}$Li) is well reproduced by the phenomenological description, in spite of the relatively high distortion factor. It indicates that the accurate prediction of the unit cross section for this case by Eq. (\ref{eq:phen}) is probably coincidental. We conclude that the validity of Eq. (\ref{eq:phen}) is uncertain for the mass range $6<A<12$.

\section{Summary and Conclusions}
By complementing available data for GT unit cross section for the ($^{3}$He,$t$) reaction at 140 $A$MeV with existing, new, and reevaluated data for the ($t$,$^{3}$He) reaction at 115 $A$MeV, a systematic picture of the mass dependence of the GT unit cross section was achieved for these reactions in the target mass range of $1<A<120$. The small difference in beam energy between the two probes does not noticeably (i.e. within statistical and systematic error margins) affect the GT unit cross sections, given the consistency for the extracted unit cross section from the two probes in the overlapping mass region. For both probes and for target masses with $A\geq12$, the GT unit cross sections are well described by a simple function (Eq. (\ref{eq:phen})). In the analysis of ($t$,$^{3}$He) and ($^{3}$He,$t$) charge-exchange data, this simple function can directly be used to extract GT strengths directly from experimental differential cross sections.

The components that make up the unit cross section in eikonal approximation (a kinematic factor $K$, the distortion factor $N^{D}$ and the volume integral of the effective ${\sigma\tau}$ operator $|J_{\sigma\tau}|$) can also be described by simple functions of mass number $A$ and by combining these equations a description equal in quality to the use of Eq. (\ref{eq:phen}) for target masses $A\geq12$  is achieved. In addition, the availability of a separate equation for the mass dependence of  $|J_{\sigma\tau}|$ allows for the calculation of the unit cross sections for $A<12$, where distortion factors must be calculated on a case-by-case basis.

Although there is a rough qualitative correspondence, the extracted values of $|J_{\sigma\tau}|$ from the data are systematically lower (on average by about 20\%) than the values predicted in Born Approximation if a short-range approximation for exchange contributions to the transition amplitude is used. This discrepancy is consistent with the findings of earlier works \cite{UDA87,KIM00} in which the exact treatment of exchange terms for composite probes was compared with the short-range approximation. To make further progress, a general tool to calculate charge-exchange reactions involving composite probes that treats exchange contributions exactly is needed.

\section{Acknowledgements}
We thank the NSCL staff for their support during the ($t$,$^{3}$He) experiments presented in this paper. This work was supported by the US NSF (PHY-0822648 (JINA)and PHY-0606007). R.Z. wishes to thank Takeshi Udagawa, Yoshitaka Fujita, Tatsuya Adachi, Takashi Nakamura, Masaki Sasano, Kentaro Yako, Dieter Frekers, Vladimir Zelevinsky, Alex Brown and Filomena Nunes for fruitful and stimulating  discussions concerning various aspects of this paper.

\bibliography{prc}

\end{document}